\documentstyle[prb,twocolumn,aps]{revtex}
\input{epsf}
\begin{document}

\tolerance 10000

\twocolumn[\hsize\textwidth\columnwidth\hsize\csname %
@twocolumnfalse\endcsname

\draft

\title{Spectroscopy of Matter Near Criticality}

\author{B. A. Bernevig, D. Giuliano, and R. B. Laughlin}

\address{Department of Physics, Stanford University,
        Stanford, California 94305}

\date{\today}
\maketitle
\widetext

\begin{abstract}
\begin{center}
\parbox{14cm}{We propose that the finite-frequency susceptibility
              of matter near a class of zero-temperature phase transition
              exhibits distinctive excitonic structure similar to
              meson resonances.}

\end{center}
\end{abstract}

\pacs{
\hspace{1.9cm}
}
]

\narrowtext

In this paper we predict a new spectroscopic effect that should occur very
generally at quantum phase transitions described by $O(n)$ $\sigma$-models
\cite{sigma}

\begin{equation}
{\cal L} = | \partial_\nu \vec{\sigma} |^2 + \mu |\vec{\sigma}|^2 -
\lambda | \vec{\sigma}|^4 \; \; \; .
\end{equation}

\noindent
These are thought to describe any continuous phase transition from a
fully gapped quantum disordered insulator to a state with continuous
broken symmetry, such as an antiferromagnet or a superconductor.  The
effect we predict, shown in Fig. 1, is a sequence of resonances in the
susceptibility

\begin{equation}
\chi_q (\omega) =
\int \int \vec{\sigma}(\vec{r},t) \cdot \vec{\sigma}( 0 , 0) \;
e^{i ( q \cdot r - \omega t)} \; dr dt
\; \; \; ,
\end{equation}

\noindent
when the system is slightly detuned from criticality.  On the
insulating side these may be understood physically as bound states of
excitons.  On the ordered side they may be understood as bound states
of Goldstones.

\begin{figure}
\epsfbox{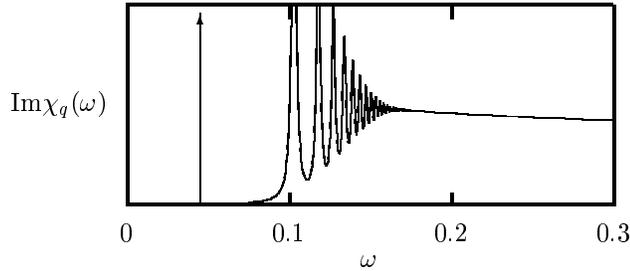}
\caption{Proposed spin susceptibility of $\sigma$-model at $q = 0$
         slightly detuned from its transition on the disordered side.}
\end{figure}

This effect is important because it is formally similar to a meson
spectrum in particle physics, and thus relevant to the larger issue of
whether behavior like that of the empty vacuum of space might occur as an
emergent phenomenon in ordinary matter.  It has been known since the 1970s
that matter undergoing a continuous zero-temperature phase transition is a
serious candidate for this, and there is widespread agreement that the
low-energy properties of such matter should be described by simple
relativistic field theories \cite{wilson}.  However, little is known
beyond these basic facts. First-principles theory is impossibly difficult
except in a few limiting cases, and all well-characterized quantum phase
transitions presently known in the laboratory are either first-order, such
as the solidification of $^4$He under pressure, or disorder-dominated.
Thus the entire subject of matter undergoing a continuous quantum phase
transition in the absence of disorder is largely unexplored,
notwithstanding its larger implications \cite{danny}.

\begin{figure}
\epsfbox{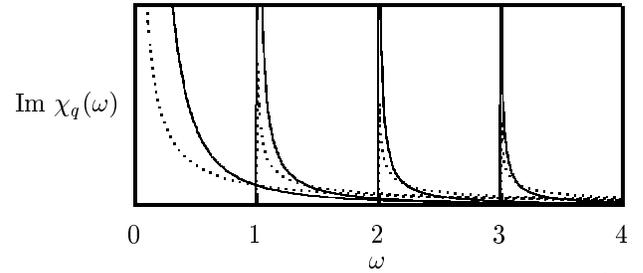}
\caption{Imaginary part of critical susceptibility $\chi_q (\omega)$
         of the O(3) $\sigma$ model in 2 spatial dimensions as given by
         Eq. (3) for equally-spaced values of $q$.  The dashed
         curves are the imaginary part of the spin chain susceptibility
         given by Eq. (4).  The limit of small $q$ is assumed
         in both cases.}
\end{figure}

Our prediction rests on a physical analogy between critical systems and
1-dimensional spin chains.  The standard O(3) $\sigma$-model has a quantum
phase transition as a function of $\mu$.  The susceptibility at the phase
transition has been calculated by $\epsilon$-expansion by various groups
\cite{groups} and found to have the relativistic form $q^{\eta-2}$, or

\begin{equation}
\chi_q (\omega) \sim \frac{1}
{[ v^2  q^2 - \omega^2 ]^{1 - \eta/2}}
\; \; \; ,
\end{equation}

\noindent
where $v$ is the speed characterizing the Goldstone mode on the ordered
side of the transition and $\eta \simeq 0.031 \pm 0.022$ is the anomalous
exponent \cite{itzykson}.  Half-integral spin chains {\it not} at a phase
transition, on the other hand, have been calculated by exact
diagonalization methods to have the strikingly similar susceptibility
\cite{haldane}

\begin{equation}
\chi_q (\omega) \sim \frac{1}
{[ v^2  (q - \pi)^2 - \omega^2 ]^{1/2}}
\; \; \; ,
\end{equation}

\noindent
where $v$ is a parameter characteristic of antiferromagnetic exchange.  As
shown in Fig. 2, in either case one sees a diverging branch cut with a
threshold on a relativistic light line.  Our idea is that that these
spectra are similar because they are the physically the same thing. The
spin chain spectra are broad because the lowered dimensionality
effectively makes the system critical by preventing antiferromagnetism and
destabilizing the system toward spin-peirels order. The higher-dimensional
critical points have a ground state and excitation spectrum similar to
that of the spin chains. However, it has been known since the work of
Fadeev and Takhtajan \cite{fadeev} that the breadth of the spin-chain
spectra is due ultimately to decay of the injected spin wave into two or
more {\it spinons} - neutral, spin-1/2 particles not postulated in the
underlying equations of motion that may be separated to infinity, do not
bind, and do not decay.  Thus our idea is that critical points in higher
dimension possess new, previously unidentified elementary excitations with
integrity analogous to spinons.  The pairwise binding of these away from
criticality is the basis of the prediction in Fig. 1. 

Spinons are particularly easy to write down and explain in the case of the
Haldane-Shastry Hamiltonian \cite{shastry}

\begin{equation}
{\cal H}_{HS} = J (\frac{2\pi}{N})^2 \sum_{\alpha < \beta}^N
\frac{\vec{S}_\alpha \cdot \vec{S}_\beta}{|z_\alpha - z_\beta|^2}
\; \; \; .
\end{equation}
 
\noindent
This choice of Hamiltonian is convenient, but not essential, as spinons
are believed to be a universal feature of such models, and indeed were
originally discovered in the Bethe solution of the Heisenberg chain
\cite{fadeev,bethe}. Here $z_\alpha = \exp (i 2 \pi \alpha / N)$ is a
lattice site on the unit circle expressed as a complex number, and
$\vec{S}_\alpha$ is a spin-1/2 spin operator for that site.  If the number
of sites $N$ is even, then the ground state of this Hamiltonian is the
spin singlet \cite{chia}

\begin{equation}
\Psi (z_1 , ... , z_{N/2}) = \prod_{j < k}^{N/2} (z_j - z_k )^2
\prod_j^{N/2} z_j \; \; \; ,
\label{ground}
\end{equation}

\noindent
where $z_j$ denotes the location of the j$^{th}$ $\uparrow$ site, all
other being down.  The energy eigenvalue of this state is $-J \pi^2(N +
5/N)/24$. If $N$ is odd, on the other hand, then the ground state is a
doublet, the $\downarrow$ component of which is one of the states

\begin{equation}
\Psi_m (z_1 , ... , z_M) = \sum_\alpha (z_\alpha^*)^m
\Phi_\alpha (z_1 , ... , z_M) 
\end{equation}

\begin{equation}
\Phi_\alpha (z_1 , ... , z_M) = 
\prod_j^M (z_\alpha - z_j) \prod_{j < k}^M (z_j - z_k )^2
\prod_j^M z_j \; \; \; ,
\end{equation}

\noindent
where $M = (N-1)/2$. The states $\Psi_m$ are propagating spinons of
momentum

\begin{equation}
q_m = N \pi / 2 - 2 \pi (m + 1/4)/N \pmod{2 \pi} \; \; \; .
\end{equation}

\noindent
They are eigenstates of the Hamiltonian with eigenvalue

\begin{equation}
E_m = - J (\frac{\pi^2}{24} )(N - \frac{1}{N} ) + \frac{J}{2}
(\frac{2\pi}{N})^2 m (M - m ) \; \; \; .
\end{equation}

\noindent
Thus the spinon may be thought of as a propagating spin-1/2 defect caused
by the addition of an extra site to an even-N chain. Combining the
momentum and energy expressions, one obtains the dispersion relation

\begin{equation}
E_q = \frac{J}{2} \biggl[ (\frac{\pi}{2})^2 - q^2 \biggr] \pmod{\pi}
\; \; \; .
\end{equation}

\noindent
This tells us that spinons are relativistic particles with Dirac points at
$q = \pm \pi / 2$.  The states $\Phi_\alpha$, which are mathematically
equivalent to fractional quantum Hall quasiparticles, describe a
$\downarrow$ spinon at site $\alpha$.

Spinons attract strongly.  The spin susceptibility

\begin{equation}
{\rm Im} \; \chi_q (\omega) =
\sum_x | < \! x | S_{q - \pi}^z | 0 \! > |^2 \; \delta ( \omega - E_x )
\end{equation}

\noindent
of the even-N Haldane-Shastry model, where $| 0 \! >$ denotes the ground
state, $| x \! >$ and $E_x$ denote an excited state and corresponding
excitation energy, and

\begin{equation}
S_{q}^z = \sum_\alpha z_\alpha^{qN/2\pi} \; S_\alpha^z
\; \; \; ,
\end{equation}

\noindent
is given exactly by \cite{haldane}

\begin{displaymath}
{\rm Im} \; \chi_{q} (\omega) = \frac{J}{8 \pi}
\int_{-\pi/2}^{\pi/2} \int_{-\pi/2}^{\pi/2} \;
\frac{|k - k'|}{\sqrt{E_k E_{k'}}}
\end{displaymath}

\begin{equation}
\times \delta(k + k' + q ) \; \delta ( E_k + E_{k'} - \omega )
\; dk \; dk'
\label{exact}
\end{equation}

\noindent
for $-\pi < q < \pi$ mod $(2\pi)$. The physical meaning of this
formula is that spinon pairs of total momentum $q$ excited out of
the vacuum escape to infinity but also attract each other, in
that the wavefunction amplitude for the two particles to coincide
is enhanced over the flat 2-spinon joint density of states by a matrix
element that strongly favors low-energy decays. This effect is crucial for
producing the divergence of the branch cut, for the joint density of
states is flat.

The strong interaction among spinons cause their field theory to be
ambiguous.  Writing the Dirac sea of electrons as

\begin{equation}
| \Phi \! > =  \prod_{|k|< \pi/2 , \sigma}
\! c_{k\sigma}^\dagger | 0 \! >  \pmod{2\pi}
\; \; \; ,
\end{equation}

\noindent
where $c_{k\sigma} = N^{-1/2} \sum_\alpha z_\alpha^{kN/2\pi} c_{\alpha
\sigma}$, we find that the even-N ground state of Eq.  (\ref{ground}) is
exactly $| \Psi \! > = P_G | \Phi \! >$, where

\begin{equation}
P_G = \prod_\alpha ( 1 - c_{\alpha \uparrow}^\dagger c_{\alpha
\downarrow}^\dagger c_{\alpha \downarrow} c_{\alpha \uparrow} )
\label{gutz}
\end{equation}

\noindent
is the Gutzwiller projector, and that the spinon-pair eigenstates are
superpositions of the wavefunctions \cite{shastry,chia}

\begin{equation}
P_G \; c_{\pi - k \uparrow}^\dagger c_{k ' \downarrow} | \Phi \! > =
P_G \; c_{\pi - k' \uparrow}^\dagger c_{k \downarrow} | \Phi \! >
\; \; \; .
\label{pair}
\end{equation}

\noindent
Thus spinons have a natural fermi representation in terms of conventional
particle and hole excitations of a Dirac sea or metal.  However, the
representation is degenerate, in that an $\uparrow$ spinon has allowed
momentum only in the range $- \pi / 2 < k < \pi / 2$ may be written either
as an $\uparrow$ particle in this range or a $\downarrow$ hole at $\pi -
k$. They are thus like Kogut-Susskind fermions, except without the
``doubling problem'' \cite{kogut}.

In order to talk concretely about spinon-like excitations potentially
present at higher dimensional critical points let us now identify a
specific transition that the $\sigma$-model might describe.  We propose
the Hamiltonian

\begin{equation}
{\cal H} = \sum_{j k \sigma} t_{jk} c_{j \sigma}^\dagger c_{k \sigma}
+ U \sum_{j \sigma \sigma '} c_{j\sigma}^\dagger c_{j\sigma '}^\dagger
c_{j \sigma '} c_{j \sigma} 
\label{hubbard}
\end{equation}

\begin{equation}
t_{jk} = |t_{jk}| \exp( i \int_j^k \vec{A} \cdot d\vec{s} )
\; \; \; \; \; \; \; \;
\oint_{\rm plaquette} \! \! \! \! \! \! \! \! \! \!
\vec{A} \cdot d\vec{s} = \pi
\end{equation}

\begin{figure}
\epsfbox{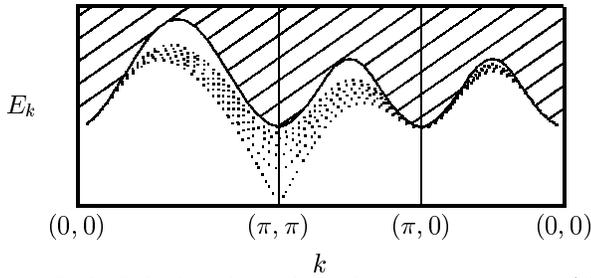}
\caption{Spin-1 exciton dispersion defined by Eqs. (4) - (7) for the
         case of $m_0 = 0.4$ for various values of $U$ approaching the
         critical value.  The solid line is the pair-production
         threshold.}
\end{figure}

\noindent
borrowed from cuprate superconductor theory \cite{zou}.  This
describes electrons moving on a square planar lattice in the presence of a
magnetic flux $\pi$ per plaquette, near-neighbor and second-neighbor
integrals $t$ and $t'$, and an on-site coulomb repulsion $U$.  When $U$ is
small this system is a conventional insulator with band structure

\begin{equation}
E_k = \pm 2 t \sqrt{ \cos^2(k_x) + \cos^2(k_y) + m^2
} \; \; \; ,
\label{spinon}
\end{equation}

\noindent
in Landau gauge $(\vec{A} = \pi y \hat{x})$ with $m = m_0 \sin(k_x) 
\sin(k_y)$, $m_0 = 2t'/t$.  When $U$ is large it is an ordered
antiferromagnet with Heisenberg exchange.  The magnetic field in this
model is simply a device for constructing a band insulator with one
electron per unit cell, and thus enabling the transition to be to a
half-integral antiferromagnet.

The parameter $\vec{\sigma}$ in Eq. (1) corresponds to conventional spin,
and $\mu$ corresponds roughly to $U$.  When $U$ is small, the ladder
approximation for the the spin susceptibility is appropriate and gives

\begin{equation}
\chi_q (\omega) = \frac{\chi_q^0 (\omega)}
{1 + U \chi_q^0 (\omega )} \; \; \; ,
\label{rpa}
\end{equation}

\begin{equation}
\chi_q^0 (\omega) = \frac{1}{(2\pi)^3} \int \int
{\rm Tr} [ G_k (E) G_{k+q}(E + \omega) ] \; dE dk
\end{equation}

\begin{equation}
G_k^{-1} (E) = E - 2 t \biggl[ \cos(k_x) \alpha_x
+ \cos(k_y) \alpha_y
+ m \beta \biggr]
\end{equation}

\begin{equation}
\alpha_x =
\left[ \begin{array}{cc} 1 & 0 \\ 0 & \! \! -1 \end{array} \right]
\; \; \;
\alpha_y =
\left[ \begin{array}{cc} 0 & 1 \\ 1 & 0 \end{array} \right]
\; \; \;
\beta =
\left[ \begin{array}{cc} 0 & \! \! -i \\ i & 0 \end{array} \right]
\; \; \; .
\end{equation}

\noindent
The imaginary part of $\chi_q (\omega)$ consists of a broad continuum with
an edge at the pair-creation threshold and a $\delta$-function
representing a spin-1 exciton bound down below this threshold by an amount
depending on $U$. This is illustrated in Fig. 3.  This exciton at momentum
$Q = (\pi, \pi)$ has integrity, i.e. does not decay, even when this
calculation is corrected to all orders, because it is the lowest-energy
excitation of the system.  The energy gap at $(\pi,\pi)$ decreases as $U$
is increased and collapses to zero at the critical value $U_c$.  As it
does so the exciton dispersion becomes relativistic.  At the transition
the Hartree-Fock equations become unstable to spin density wave formation,
and the same ladder sum computed with the broken-symmetry electron
propagator then gives three relativistic $\delta$-functions - two
light-like spin waves and one massive amplitude mode.  The mass of the
latter converges to zero at the transition.  Both the integrity of the
spin waves and their linear dispersion survive correction to all orders by
virtue of Goldstone's theorem \cite{goldstone}. 

Let us now guess, by analogy with Eq. (\ref{pair}) that the spinons that
emerge at the transition are Gutzwiller projections of conventional
particles and holes of this band structure with $m_0$ set to zero. Note
that this is {\it not} equivalent to assuming that the true electron gap
collapses.  The system continues to be an insulator across the transition
and has no low-energy charged excitations of any kind.  This makes the
spinons properly relativistic, forces their momenta to add up properly to
that of the corresponding spin wave, restricts their momenta to half the
Brillouin zone, and causes an $\uparrow$ particle at momentum k and a
$\downarrow$ hole a momentum $(\pi , \pi ) -k$ to describe the same
spinon. 

Let us now simulate the attraction between spinons by re-evaluating Eq. 
(\ref{rpa}) with $m_0$ set to zero and $U$ tuned to criticality.  This is
done in Fig. 4.  It may be seen that one properly obtains a continuum with
a divergent branch cut on the relativistic light line.  As is the case in
1 dimension, the spectrum is a continuum because the spinon pairs always
escape to infinity.  The threshold is sharp because lower-energy decays
are kinematically forbidden.  The edge is enhanced because the attractive
force between spinons enhances the wavefunction when the particles come
together with small relative momentum.  But the enhancement is divergent
in this case because of criticality.  At the phase transition an
arbitrarily small perturbation must be able to push the system into the
ordered state, i.e.  cause the spinons to bind. This forces the
susceptibility to be infinite. The exponent of the divergence is 1 in this
calculation rather than the correct $2 - \eta$, but this simply means that
spinons interact by means of a potential with a long-range tail, an effect
we have verified numerically.

\begin{figure}
\epsfbox{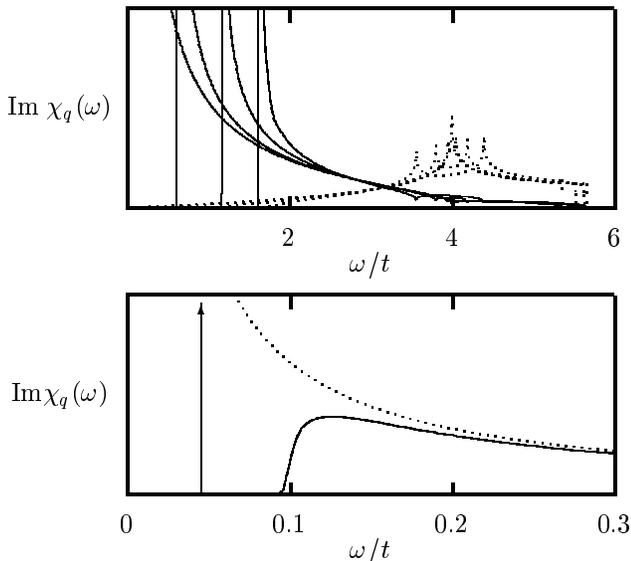}
\caption{Top: Imaginary part of the susceptibility $\chi_q (\omega)$
         defined by Eq. (4) for $m_0 = 0$ and the critical value of $U$
         for equally-spaced values of $q$ near $(\pi,\pi)$. The dashed
         curves are the imaginary part of $\chi^0_q (\omega)$. Bottom:
         The same calculation for $q = (\pi,\pi)$ and $m_0 = 0.1$. The
         dotted curve is the $m_0 \rightarrow 0$ limit.}
\end{figure}

Let us now simulate the effect of pushing to the disordered side of the
transition by increasing $m_0$ slightly keeping $U$ fixed.  As shown in
Fig. 4, this opens a gap in the continuum and allows for a single bound
state, which is the exciton in Fig. 3.  The presence of this continuum and
its matching the critical susceptibility at high frequencies is required
by the principle that a fast measurement should not be able to distinguish
a critical system from an almost-critical one.  The gap in the continuum
at low frequencies is required for the bound state to be sharp.  The sharp
bound state is required by the integrity of the exciton as the
lowest-energy excitation of the disordered phase.

However, a more thoughtful analysis leads to the spectrum of Fig. 1.  The
continuum threshold in the bottom of Fig. 4 cannot be physically right
because spinons are not legitimate elementary excitations of a band
insulator.  They must cease to exist immediately when the system is pushed
away from criticality.  The simplest way for this to occur is if the
long-range tail of the interaction diverges so as to cause confinement.  
The result would be a bound-state spectrum more like that of a meson than
a traditional exciton.  The spectrum in Fig. 1 is calculated assuming that
the interaction between the spinons is proportional to $\ln (r)$ rather
than a point interaction.  We specifically compute

\begin{equation}
{\rm Im} \; \chi_q (\omega) \sim \sum_n | \psi_n (0) |^2 \; \delta
( \omega - E_n )
\end{equation}

\begin{equation}
- \frac{1}{2} \nabla^2 \psi_n + \ln (r) \psi_n = E_n \psi_n
\; \; \; .
\end{equation}

\noindent
The other obvious possibility is a string potential \cite{diamantini},
expected on the ordered side of the transition, but the result in that
case is not much different. In either case one sees a series of bound
state resonances that merge into the continuum as the energy scale is
raised.  Whether these have integrity or decay is determined simply by
whether they are above or below the pair-production threshold for the
first resonance.  These resonances maybe understood physically as bound
states of excitons, but a more apt analogy would be with charmonium
\cite{applequist}.

Thus if these resonances are observed experimentally it will suggest
strongly that a relativistic gauge theory of some kind, most likely SU(2)
in this case \cite{baskeran}, can emerge at critical points. 

This work was supported primarily by the NSF under grant No. DMR-9813899.
Additional support was provided by NASA Collaborative Agreement NCC
2-794 and by NEDO.

\end{document}